\documentclass{article}
\input epsf

\begin{document}

\hfill{\vbox{\hbox{OSU HEP 99-02}
}}

\vskip .75in

\begin{center}
  {\LARGE The Higgs Boson Mass in Gauge-Mediated Supersymmetry-Breaking
   Models with Generalized Messenger Sectors\footnote{Submitted
   to {\it Physical Review} {\bf D}.}}
  \vskip .3in
  {\large
    Thomas A.~Kaeding\footnote{kaeding@okstate.edu} and
    Satyanarayan Nandi\footnote{shaown@okstate.edu}\\
    ~\\
    Department of Physics, Oklahoma State University\\
    Stillwater, Oklahoma~~74078
  }
  \vskip 1.5em
  {\large 1999 June 11}
  \vskip .5in
\end{center}

\begin{abstract}
The lighter neutral scalar Higgs mass is examined in gauge-mediated supersymmetry-breaking
models in which the messenger sector responsible for SUSY breaking is
allowed to involve more general sets of 
$SU(3) \otimes SU(2) \otimes U(1)$
multiplets than those contained in the SU(5) $5 + \bar{5}$ or
$10 + \bar{10}$ multiplets.
The largest mass for the lighter scalar Higgs is found to be 132 GeV when the
breaking parameter $\Lambda$ is taken to be 100 TeV\@.
Thus the predictions for the lightest Higgs mass can be tested at the
upgraded Tevatron runs for these general classes of GMSB models.
\end{abstract}

\newpage

\section{Introduction}

One of the main goals at present and the next generation of colliders
is to search for evidence of the Higgs boson of the Standard Model (SM).
Theories of supersymmetry (SUSY)
posess extended Higgs sectors with multiple Higgs bosons.
Common to all models of low energy supersymmetry where the couplings
remain perturbative up to very high energies, however, is the prediction
of a relatively low mass Higgs boson \cite{low}. Thus the search for the lightest
Higgs boson is a crucial test for SUSY theories. The question arises as
to the kinds of limits that can be placed on the mass of the lightest
Higgs boson in various SUSY models.

In this paper, we explore the values that the mass of the lightest Higgs
boson can have in theories of gauge-mediated supersymmetry-breaking
(GMSB). GMSB models provide an alternative to models where supersymmetry
breaking is communicated to the visible sector by gravitational
interactions \cite{gmsb_}.
The defining characteristic of GMSB models is that the
SUSY breaking is communicated to the visible sector by gauge
interactions. The messenger sector is composed of some set of
superfields with SM couplings, but which are not part of the spectrum
of the Minimal Supersymmetric Standard Model (MSSM).
In minimal models of GMSB, these superfields are assumed to form
complete SU(5) multiplets so that the apparent unification of the
gauge couplings can be preserved.
However, the superfields that act as messengers need not form complete
GUT multiplets. Other superfields can be present at the messenger
scale that do not act as messengers and have little or no effect
on the masses of the MSSM particles, but can still participate in
ensuring gauge coupling unification \cite{martin}.
The messenger sector in these generalized GMSB models
thereby contains superfields that transform as
$SU(3) \otimes SU(2) \otimes U(1)$ mulitplets, but not as complete
GUT multiplets. We will explore the values that the mass of the lightest
Higgs boson can acquire in minimal GMSB models and the generalized GMSB
models discussed in \cite{martin}.

We seek the answers to two questions. First, is there a bound on the
lightest Higgs mass in these general classes of
$SU(3) \otimes SU(2) \otimes U(1)$ multiplets as we vary the parameters
of the models? Second, are the predictions for the lightest Higgs mass
still within the detectable range of the upcoming Tevatron runs?

\section{Generalized Messenger Sector}

Here we will outline the generalization of the messenger sector as 
suggested by \cite{martin}.

We assume that the messenger fields couple to a single chiral superfield
$S$, so that the superpotential contains the term
\begin{equation}
  W \ni \lambda_iS\Phi_i\overline\Phi_i.
\end{equation}
The induced gaugino mass parameters are (for $a=1,2,3$)
\begin{equation}
  M_a = \frac{\alpha_a}{4\pi} \frac{F}{S} \sum_i n_{ai} g(x_i)
\end{equation}
where
\begin{equation}
  x_i = \left| F/\lambda_i S^2 \right|
\end{equation}
where $S$ and $F$ here represent the vacuum expectation values (VEVs) of the
chiral scalar superfields
$S$ and $F$, respectively.
The function $g$ is defined by
\begin{equation}
  g(x) \equiv \frac{1}{x^2} \left[(1+x) \log (1+x) + (1-x) \log (1-x)\right]
\end{equation}
and $n_{ai} = 1$ for a pair of $\overline{\bf N}+{\bf N}$ of SU($N$),
and $n_{ai} = 3$ for an antisymmetric two-index tensor of SU($N$).
We use the GUT normalization for $\alpha_1$ so that
$n_1 = \frac{6}{5}Y^2$ for a pair with hypercharge $Y = Q - T_3$.

We define the convenient numbers
\begin{equation}
  N_a \equiv \sum_i n_{ai}.
\end{equation}

The requirement that the MSSM gauge couplings should stay perturbative
($\alpha_i \leq 0.2$) up to the unification scale amounts to
\begin{equation}
  \label{Ncondition}
  \begin{array}{rcl}
    N_1 & \leq & 4,\\
    N_2 & \leq & 4,\\
    N_3 & \leq & 4.
  \end{array}
\end{equation}
We shall therefore consider messenger sectors that contain up to four
sets of SU(5) multiplets ${\bf 5} + \overline{\bf 5}$, or one
${\bf 10} + \overline{\bf 10}$, or one ${\bf 5} + \overline{\bf 5}$ together
with one ${\bf 10} + \overline{\bf 10}$.
We will also consider generalized messenger sectors consisting of
various multiplets of  SU(3) $\times$ SU(2) $\times$ U(1) as in
\cite{martin}.
Consider fields that transform under SU(3) $\times$ SU(2) $\times$ U(1) as
\begin{equation}
  \begin{array}{rcl}
    Q + \overline{Q} & = & ({\bf 3}, {\bf 2}, \frac{1}{6}) +
                           (\overline{\bf 3}, {\bf 2}, -\frac{1}{6}) \ ,\\
    U + \overline{U} & = & (\overline{\bf 3}, {\bf 1}, -\frac{2}{3}) +
                           ({\bf 3}, {\bf 1}, \frac{2}{3}) \ ,\\
    D + \overline{D} & = & (\overline{\bf 3}, {\bf 1}, \frac{1}{3}) +
                           ({\bf 3}, {\bf 1}, -\frac{1}{3}) \ ,\\
    L + \overline{L} & = & ({\bf 1}, {\bf 2}, -\frac{1}{2}) +
                           ({\bf 1}, {\bf 2}, \frac{1}{2}) \ ,\\
    E + \overline{E} & = & ({\bf 1}, {\bf 1}, 1) +
                           ({\bf 1}, {\bf 1}, -1) \ .
  \end{array}
\end{equation}
The number of sets of $Q + \overline{Q}$ fields will be denoted by $n_Q$,
the number of sets of $U + \overline{U}$ by $n_U$, etc.
The number of sets of SU(5) ${\bf 5} + \overline{\bf 5}$ will be denoted
by $n_5$, and the number of ${\bf 10} + \overline{\bf 10}$ by $n_{10}$.
Their contributions to the $N_a$ are given by
\begin{equation}
  \begin{array}{rcl}
    N_1 &=& \frac{1}{5} (n_Q + 8 n_U + 2 n_D + 3 n_L + 6 n_E)
            + n_5 + 3 n_{10},\\
    N_2 &=& 3 n_Q + n_L + n_5 + 3 n_{10},\\
    N_3 &=& 2 n_Q + n_U + n_D + n_5 + 3 n_{10}.
  \end{array}
\end{equation}
Requiring gauge coupling unification as well as gauge coupling
perturbativity, we have the requirement that
$(n_Q, n_U, n_D, n_L, n_E)$ be less than (1,0,2,1,2), or
(1,1,1,1,1), or (1,2,0,1,0), or (0,0,4,4,0).
Using these fields, the possible combinations satisfying this
condition with $N_1 \neq 0$, $N_2 \neq 0$, $N_3 \neq 0$ are
\cite{martin}:
($\frac{1}{5}$, 3, 2),
($\frac{3}{5}$, 3, 3),
($\frac{4}{5}$, 4, 2),
(1, 1, 1),
(1, 3, 4),
($\frac{6}{5}$, 4, 3),
($\frac{7}{5}$, 1, 2),
($\frac{7}{5}$, 3, 2),
($\frac{8}{5}$, 2, 1),
($\frac{8}{5}$, 4, 4),
($\frac{9}{5}$, 1, 3),
($\frac{9}{5}$, 3, 3),
(2, 2, 2),
(2, 4, 2),
($\frac{11}{5}$, 1, 1),
($\frac{11}{5}$, 1, 4),
($\frac{11}{5}$, 3, 1),
($\frac{11}{5}$, 3, 4),
($\frac{12}{5}$, 2, 3),
($\frac{12}{5}$, 4, 3),
($\frac{13}{5}$, 1, 2),
($\frac{13}{5}$, 3, 2),
($\frac{14}{5}$, 2, 4),
($\frac{14}{5}$, 4, 1),
($\frac{14}{5}$, 4, 4),
(3, 3, 3),
($\frac{16}{5}$, 4, 2),
($\frac{17}{5}$, 1, 1),
($\frac{17}{5}$, 3, 4),
($\frac{18}{5}$, 4, 3),
($\frac{19}{5}$, 1, 2),
and (4, 4, 4).

\section{Calculation of the Higgs Mass}

If we assume that all $\lambda_i$ are equal, and replace the VEVs with
$\Lambda = F/S$ and the messenger scale $M$, then
the soft
SUSY breaking gaugino and scalar masses at the messenger scale are
given by \cite{{martin},{gmsb},{spm}}
\begin{equation} \label{gmass}
\tilde M_a(M) = N_a \, g\left(\frac{\Lambda}{M}\right) \, \frac{\alpha_a(M)}{4 \pi} \,
\Lambda
\end{equation}
and
\begin{equation} \label{smass}
\tilde m^2(M) = 2\, f\left(\frac{\Lambda}{M}\right) \, \sum^3_{a = 1} \, k_a \, N_a \,
C_a \, \left (\frac{\alpha_a(M)}{4 \pi} \right )^2 \, \Lambda^2
\end{equation}
where the $\alpha_a$ are the three SM gauge couplings and $k_a =$ 1, 1
and 3/5
for SU(3), SU(2), and U(1), respectively. The $C_a$ are zero for gauge
singlets and are 4/3, 3/4 and $Y^2$ for the fundamental
representations of SU(3), SU(2) and U(1), respectively (with $Y$ given
by $Q = I_3 + Y$). The functions $g(x)$ and $f(x)$ are messenger scale threshold
functions. We calculate the sparticle masses at the messenger scale $M$
using Eqs.~(\ref{gmass}) and (\ref{smass}) and run these
to the electroweak scale using the appropriate renormalization group
equations \cite{bbo}.

The calculation of the Higgs mass from the sparticle spectrum is carried
out as in \cite{castano}, with one-loop corrections given in
\cite{haber}.
The tree-level potential for the Higgs is
\begin{equation}
  V_{\hbox{tree}} = m_1^2 H_1^\dagger H_1
                  + m_2^2 H_2^\dagger H_2
                  + m_{12}^2 \left( H_1 H_2 + H_2 H_1 \right)
                  + ...
\end{equation}
where
\begin{equation}
  \begin{array}{rcl}
    m_1^2 &=& m_{H_1}^2 + \mu^2,\\
    m_2^2 &=& m_{H_2}^2 + \mu^2,\\
    m_{12}^2 &=& B \mu.
  \end{array}
\end{equation}
The mass of the pseudoscalar Higgs is then
\begin{equation}
  M_A^2 = m_1^2 + m_2^2.
\end{equation}

We can parameterize the corrections to
the mass matrix elements for the scalar Higgs sector by
\begin{equation}
  \begin{array}{rcl}
    M_{11} &=& M_A^2 \sin^2 \beta + M_Z^2 \cos^2 \beta + (\Delta M^2_{11})_{\hbox{\scriptsize 1LL}} +
               (\Delta M^2_{11})_{\hbox{\scriptsize mix}}\\
    M_{22} &=& M_A^2 \cos^2 \beta + M_Z^2 \sin^2 \beta + (\Delta M^2_{22})_{\hbox{\scriptsize 1LL}} +
               (\Delta M^2_{22})_{\hbox{\scriptsize mix}}\\
    M_{12} = M_{21} &=& -(M_A^2 + M_Z^2) \sin \beta \cos \beta +
               (\Delta M^2_{12})_{\hbox{\scriptsize 1LL}} + (\Delta M^2_{12})_{\hbox{\scriptsize mix}}
  \end{array}
\end{equation}
For completeness, we include the expressions for the corrections.
The one-loop leading-log corrections are \cite{haber}
\begin{equation}
  \begin{array}{rcl}
    (\Delta M^2_{11})_{\hbox{\scriptsize 1LL}} &=&
      \frac{g^2 m_Z^2 \cos^2 \beta}{96 \pi^2 \cos^2 \theta_W}
      \left[P_t \ln \frac{M_Q M_U}{m_t^2}
      \right.\\&&\left.
      + \left(36 \frac{m_b^4}{m_Z^4 \cos^4 \beta}
      - 18 \frac{m_b^2}{m_Z^2 \cos^2 \beta}
      + P_b +P_f +P_g +P_{2H}\right) \ln \frac{M_Q M_D}{m_Z^2}
      \right.\\&&\left.
      + \Theta (m_{A^0} - m_Z) (P_{1H} - P_{2H}) \ln \frac{m_{A^0}^2}{m_Z^2}
      \right.\\&&\left.
      - 9 \left(1 + 4 e_b \sin^2 \theta_W\right)
      \left(\frac{m_b^2}{m_Z^2 \cos^2 \beta} - \frac{1}{6}\right)
      \ln \left(\frac{M_Q^2}{M_D^2}\right)
      \right.\\&&\left.
      + \frac{3}{2} \left(1 - 4 e_t \sin^2 \theta_W\right)
      \ln \left(\frac{M_Q^2}{M_U^2}\right)\right],
      \\
    (\Delta M^2_{22})_{\hbox{\scriptsize 1LL}} &=&
      \frac{g^2 m_Z^2 \sin^2 \beta}{96 \pi^2 \cos^2 \theta_W}
      \left[(P_b + P_f + P_g + P_{2H}) \ln \frac{M_Q M_D}{m_Z^2}
      \right.\\&&\left.
      + \left(36 \frac{m_t^4}{m_Z^4 \sin^4 \beta}
      - 18 \frac{m_t^2}{m_Z^2 \sin^2 \beta} + P_t\right) \ln \frac{M_Q M_U}{m_t^2}
      \right.\\&&\left.
      + \Theta (m_{A^0} - m_Z) (P_{1H} - P_{2H}) \ln \frac{m_{A^0}}{m_Z^2}
      \right.\\&&\left.
      + 6 \frac{m_t^2}{m_Z^2 \sin^4 \beta}
      \right.\\&&\left.
      - 9 \left(1 - 4 e_t \sin^2 \theta_W\right)
      \left(\frac{m_t^2}{m_Z^2 \sin^2 \beta} - \frac{1}{6}\right)
      \ln \left(\frac{M_Q^2}{M_U^2}\right)
      \right.\\&&\left.
      + \frac{3}{2} \left(1 + 4 e_b \sin^2 \theta_W\right)
      \ln \left(\frac{M_Q^2}{M_D^2}\right)\right],
      \\
    (\Delta M^2_{12})_{\hbox{\scriptsize 1LL}} = (\Delta M^2_{21})_{\hbox{\scriptsize 1LL}} &=&
      - \frac{g^2 m_Z^2 \sin \beta \cos \beta}{96 \pi^2 \cos^2 \theta_W}
      \left[\left(P_t - 9 \frac{m_t^2}{m_Z^2 \sin^2 \beta}\right) \ln \frac{M_Q M_U}{m_t^2}
      \right.\\&&\left.
      + \left(P_b - 9 \frac{m_b^2}{m_Z^2 \cos^2 \beta} + P_f + P_g' + P_{2H}'\right)
      \ln \frac{M_Q M_D}{m_Z^2}
      \right.\\&&\left.
      - \Theta (m_{A^0} - m_Z) (P_{1H} - P_{2H}') \ln \frac{m_{A^0}}{m_Z^2}
      \right.\\&&\left.
      - \frac{9}{2} \left(1 - 4 e_t \sin^2 \theta_W\right)
      \left(\frac{m_t^2}{m_Z^2 \sin^2 \beta} - \frac{1}{3}\right)
      \ln \left(\frac{M_Q^2}{M_U^2}\right)
      \right.\\&&\left.
      - \frac{9}{2} \left(1 + 4 e_b \sin^2 \theta_W\right)
      \left(\frac{m_b^2}{m_Z^2 \cos^2 \beta} - \frac{1}{3}\right)
      \ln \left(\frac{M_Q^2}{M_D^2}\right)\right],
  \end{array}
\end{equation}
where $M_U$, $M_D$, and $M_Q$ refer to the soft mass parameters and
\begin{equation}
  \begin{array}{rcl}
    P_t & = & 3 (1 - 4 e_t \sin^2 \theta_W + 8 e_t^2 \sin^4 \theta_W),\\
    P_b & = & 3 (1 + 4 e_b \sin^2 \theta_W + 8 e_b^2 \sin^4 \theta_W),\\
    P_f & = & 6 [3 - 6 \sin^2 \theta_W + 4 (1 + 2 e_t^2 + 2 e_b^2) \sin^4 \theta_W],\\
    P_g & = & -44 + 106 \sin^2 \theta_W - 62 \sin^4 \theta_W,\\
    P_g' & = & 10 + 34 \sin^2 \theta_W - 26 \sin^4 \theta_W,\\
    P_{2H} & = & -10 + 2 \sin^2 \theta_W - 2 \sin^4 \theta_W,\\
    P_{2H}' & = & 8 - 22 \sin^2 \theta_W + 10 \sin^4 \theta_W,\\
    P_{1H} & = & -9 \cos^4 2\beta + (1 - 2 \sin^2 \theta_W + 2 \sin^4 \theta_W) \cos^2 2\beta,\\
  \end{array}
\end{equation}
and the quark charges are $e_t$ = 2/3 and $e_b$ = $-$1/3.
The contributions due to squark mixing are \cite{haber}
\begin{equation}
  \label{squarkmix}
  \begin{array}{rcl}
    (\Delta M^2_{11})_{\hbox{\scriptsize mix}} &=&
      \frac{3 g^2}{16 \pi^2 m_W^2}
      \left\{\frac{m_b^4A_bX_b}{\cos^2\beta}\left[2h(M_Q^2,M_D^2) + A_bX_bg(M_Q^2,M_D^2)\right]
      \right.\\&&\left.
      + \frac{m_t^4\mu^2X_t^2}{\sin^2\beta}g(M_Q^2,M_U^2)
      \right.\\&&\left.
      + m_Z^2m_t^2\mu\cot\beta[X_tp_t(M_Q^2,M_U^2)-\mu\cot\beta B(M_Q^2,M_U^2)]
      \right.\\&&\left.
      + m_Z^2m_b^2A_b[X_bp_b(M_Q^2,M_D^2)-A_bB(M_Q^2,M_D^2)]\right\},\\
    (\Delta M^2_{22})_{\hbox{\scriptsize mix}} &=&
      \frac{3 g^2}{16 \pi^2 m_W^2}
      \left\{\frac{m_t^4A_tX_t}{\sin^2\beta}\left[2h(M_Q^2,M_U^2) + A_tX_tg(M_Q^2,M_U^2)\right]
      \right.\\&&\left.
      + \frac{m_b^4\mu^2X_b^2}{\cos^2\beta}g(M_Q^2,M_D^2)
      \right.\\&&\left.
      + m_Z^2m_b^2\mu\tan\beta[X_bp_b(M_Q^2,M_D^2)-\mu\tan\beta B(M_Q^2,M_D^2)]
      \right.\\&&\left.
      + m_Z^2m_t^2A_t[X_tp_t(M_Q^2,M_U^2)-A_tB(M_Q^2,M_U^2)]\right\},\\
    (\Delta M^2_{12})_{\hbox{\scriptsize mix}} = (\Delta M^2_{21})_{\hbox{\scriptsize mix}} &=&
      -\frac{3 g^2}{32 \pi^2 m_W^2}
      \left\{\frac{2m_t^4}{\sin^2\beta}\mu X_t[h(M_Q^2,M_U^2) + A_tX_tg(M_Q^2,M_U^2)]
      \right.\\&&\left.
      + \frac{2m_b^4}{\cos^2\beta}\mu X_b[h(M_Q^2,M_D^2) + A_bX_bg(M_Q^2,M_D^2)]
      \right.\\&&\left.
      - m_Z^2m_b^2\tan\beta[(\mu^2+A_b^2)B(M_Q^2,M_D^2) - X_bY_bp_b(M_Q^2,M_D^2)]
      \right.\\&&\left.
      - m_Z^2m_t^2\cot\beta[(\mu^2+A_t^2)B(M_Q^2,M_U^2) - X_tY_tp_t(M_Q^2,M_U^2)]\right\}
  \end{array}
\end{equation}
where
\begin{equation}
  \begin{array}{rcl}
    X_t & = & A_t - \mu \cot \beta,\\
    X_b & = & A_b - \mu \tan \beta,\\
    Y_t & = & A_t + \mu \tan \beta,\\
    Y_b & = & A_b + \mu \cot \beta,
  \end{array}
\end{equation}
and the functions $h$, $B$, $p_b$, and $p_t$ are
\begin{equation}
  \begin{array}{rcl}
    h (a, b) & = & \frac{1}{a-b} \ln \frac{a}{b},\\
    B (a, b) & = & \frac{1}{(a-b)^2} \left[\frac{1}{2} (a + b)
      - \frac{ab}{a-b} \ln \frac{a}{b}\right],\\
    p_b (a, b) & = & f (a, b) - 2 e_b \sin^2 \theta_W (a-b) g (a, b),\\
    p_t (a, b) & = & f (a, b) + 2 e_t \sin^2 \theta_W (a-b) g (a, b),\\
    g (a, b) & = & \frac{1}{(a-b)^2} \left[2 - \frac{a+b}{a-b} \ln \frac{a}{b}\right],\\
    f (a, b) & = & \frac{-1}{a-b} \left[1-\frac{b}{a-b} \ln \frac{a}{b}\right].
  \end{array}
\end{equation}

\section{Results}

For the various choices of $N_1$, $N_2$, and $N_3$,
the parameter space of $\Lambda$ and $\tan \beta$ was explored.
The messenger scale was fixed at $M = 10^5$ TeV\@.
Figure 1 shows two examples, for $N_1 = 1$, $N_2 = 1$, $N_3 = 1$,
and for $N_1 = 1$, $N_2 = 3$, $N_3 = 4$.
The contours of $M_h$ rise sharply for small $\tan \beta$.
They appear to
level out for large $\tan \beta$, but actually begin to turn upward again.
A large part of the region considered is excluded for having a negative
mass squared for the scalar tau lepton.
Such a mass would give a VEV that violates both electric charge and
lepton number, and is hence phenomenologically unacceptable.

Using $M = 10^5$ TeV, and considering $\Lambda$ up to 100 TeV,
we find that the largest value for the lighter scalar Higgs mass is
132 GeV\@.
The largest values for the mass in the various models with $\Lambda = 100$ TeV
are given in Table 1.
The largest mass typically occurs at $\tan \beta = 29$, with little variation on
the values of $N_a$.
The maximum value occurs at the maximum value of $\Lambda$, and would be
larger if $\Lambda$ were allowed to vary beyond 100 TeV\@.
For example, for the model with $N_1 = N_2 = N_3 = 1$
and $\Lambda$ = 200 TeV, $M_h$ = 127 GeV\@.
For the model with $N_1 = N_2 = N_3 = 4$
and $\Lambda$ = 200 TeV, $M_h$ = 139 GeV\@.
Notice also that the greatest dependence is on $N_3$,
due to the large corrections from squark mixing (Eq.~\ref{squarkmix}).

It is possible to find an analytic expression approximating the largest mass
for the lighter scalar Higgs when $\Lambda$ is fixed.
The strongest dependence is on $N_3$ and is logarithmic.
The variation with respect to $N_1$ and $N_2$ is weaker and
can be approximated by linear functions.
If we parameterize the dependence of $M_h$ on the $N_a$ by
\begin{equation}
  M_h = \left(a + b N_1 + c N_2 + d \ln N_3\right) ({\rm GeV}),
  \label{abcdeq}
\end{equation}
then for $\Lambda$ = 100 TeV, we have approximately
\begin{equation}
  a = 118.3, b = 0.1, c = 0.2, \hbox{ and} \; d = 9.1.
\end{equation}
For comparison, the values of these coefficients
for $\Lambda$ = 50 TeV and 200 TeV
are given in Table 2. 
The dependence on $N_3$ is largest, as reflected in the values of $d$.
This large, logarithmic dependence is due to the large corrections
due to squark mixing (Eq.~\ref{squarkmix}).
This dependence decreases as $\Lambda$ increases, while the constant term
$a$ also increases.

\section{Conclusion}

In this work, we have made a detailed investigation of the mass of
the lightest MSSM Higgs boson in generalized GMSB models.
We have considered a variety of messenger sectors belonging to
different vectorlike representations of the SM gauge group,
$SU(3) \otimes SU(2) \otimes U(1)$, as well as SU(5).
We imposed the condition of peturbativity on the gauge couplings
as well as the possibility of unification.
For each choice of messenger sector, we studied the variation
in the Higgs mass with respect to the GMSB parameters
$\tan \beta$ and $\Lambda$. The Higgs mass depends sensitively
on $\tan \beta$ and $\Lambda$, but not much on the messenger 
sector scale, $M$. For $\Lambda$ less than 100 TeV, the bound on
the lightest Higgs mass is 132 GeV. For $\Lambda$ equal to 500 Tev,
the bound increases to 139 GeV. These bounds are within reach of the
forthcoming upgraded Tevatron search. Thus, we conclude that the
Tevatron Run 2 and Run 3 searches will be able to explore the
lightest Higgs boson for a wide variety of messenger sector
content in GMSB models.

\section*{Acknowledgments}

We wish to thank David J. Muller for many useful discussions
during this work, specially for his help with the program. This
work is supported in part by the U.S. Department of Energy, Grant
Number DE-FG03-98ER41076.

\begin{table}
  \label{resultstable}
  \caption{Numerical results for the mass of the lighter neutral scalar Higgs boson.
           Values are for $M = 10^5$ TeV and maximum $\Lambda$ = 100 TeV\@.
           The largest mass occurs typically near $\tan\beta = 29$ and $\Lambda = 100$ GeV,
           except for the value marked with $^*$, for which the region of negative stau
           mass squared covers that point of parameter space.}
  \begin{center}
    \begin{tabular}{|l|ccc|c|}
      \hline
      Messenger Sector & $N_1$ & $N_2$ & $N_3$ & $M_h$ (GeV)\\
      \hline
      {\bf 5} $\oplus$ $\overline{\bf 5}$ & 1 & 1 & 1 & 118.5 \\
      2({\bf 5} $\oplus$ $\overline{\bf 5})$ & 2 & 2 & 2 & 125.1 \\
      3({\bf 5} $\oplus$ $\overline{\bf 5})$ & 3 & 3 & 3 & 129.0 \\
      4({\bf 5} $\oplus$ $\overline{\bf 5})$ & 4 & 4 & 4 & 131.7 \\
      {\bf 10} $\oplus$ $\overline{\bf 10}$ & 3 & 3 & 3 & 129.0 \\
      {\bf 5} $\oplus$ $\overline{\bf 5}$ $\oplus$
        {\bf 10} $\oplus$ $\overline{\bf 10}$ & 4 & 4 & 4 & 131.7 \\
      \hline
      & 1/5 & 3 & 2 & 125.1 \\
      & 3/5 & 3 & 3 & 128.9 \\
      & 4/5 & 4 & 2 & 125.5 \\
      & 1 & 1 & 1 & 118.5 \\
      & 1 & 3 & 4 & 131.5 \\
      & 6/5 & 4 & 3 & 129.1 \\
      & 7/5 & 1 & 2 & 124.8 \\
      & 7/5 & 3 & 2 & 125.3 \\
      & 8/5 & 2 & 1 & 119.0 \\
      & 8/5 & 4 & 4 & 131.7 \\
      & 9/5 & 1 & 3 & 128.6 \\
      & 9/5 & 3 & 3 & 128.9 \\
      & 2 & 2 & 2 & 125.1 \\
      & 2 & 4 & 2 & 125.5 \\
      & 11/5 & 1 & 1 & 118.6 \\
      & 11/5 & 1 & 4 & 131.3 \\
      & 11/5 & 3 & 1 & 119.3 \\
      & 11/5 & 3 & 4 & 131.6 \\
      & 12/5 & 2 & 3 & 128.8 \\
      & 12/5 & 4 & 3 & 129.1 \\
      & 13/5 & 1 & 2 & 124.9 \\
      & 13/5 & 3 & 2 & 125.3 \\
      & 14/5 & 2 & 4 & 131.5 \\
      & 14/5 & 4 & 1 & 108.6$^*$ \\
      & 14/5 & 4 & 4 & 131.7 \\
      & 3 & 3 & 3 & 129.0 \\
      & 16/5 & 4 & 2 & 125.6 \\
      & 17/5 & 1 & 1 & 118.7 \\
      & 17/5 & 3 & 4 & 131.6 \\
      & 18/5 & 4 & 3 & 129.1 \\
      & 19/5 & 1 & 2 & 124.9 \\
      & 4 & 4 & 4 & 131.7 \\
      \hline
    \end{tabular}
  \end{center}
\end{table}

\begin{table}
  \label{abcdtable}
  \caption{Values for the coefficients in Eq.~\ref{abcdeq}
           for different values of $\Lambda$.}
  \begin{center}
    \begin{tabular}{|c|cccc|}
      \hline
      $\Lambda$ (TeV) & $a$ & $b$ & $c$ & $d$ \\
      \hline
      50 & 108.9 & 0.1 & 0.3 & 9.7 \\
      100 & 118.3 & 0.1 & 0.2 & 9.1 \\
      200 & 127.0 & 0.02 & 0.2 & 8.4 \\
      \hline
    \end{tabular}
  \end{center}
\end{table}

\begin{figure}
  \caption{Contours of lighter neutral scalar Higgs mass (in GeV) in the
      $\Lambda$-$\tan\beta$ parameter space
      with $M = 10^5$ TeV\@.
      The shaded region is the area where the stau mass squared is negative.
      (a) For the case $N_1=1$, $N_2=1$, $N_3=1$.
      (b) For the case $N_1=1$, $N_2=3$, $N_3=4$.}
  \vskip 0.25in
  \epsfbox[0 0 504 300]{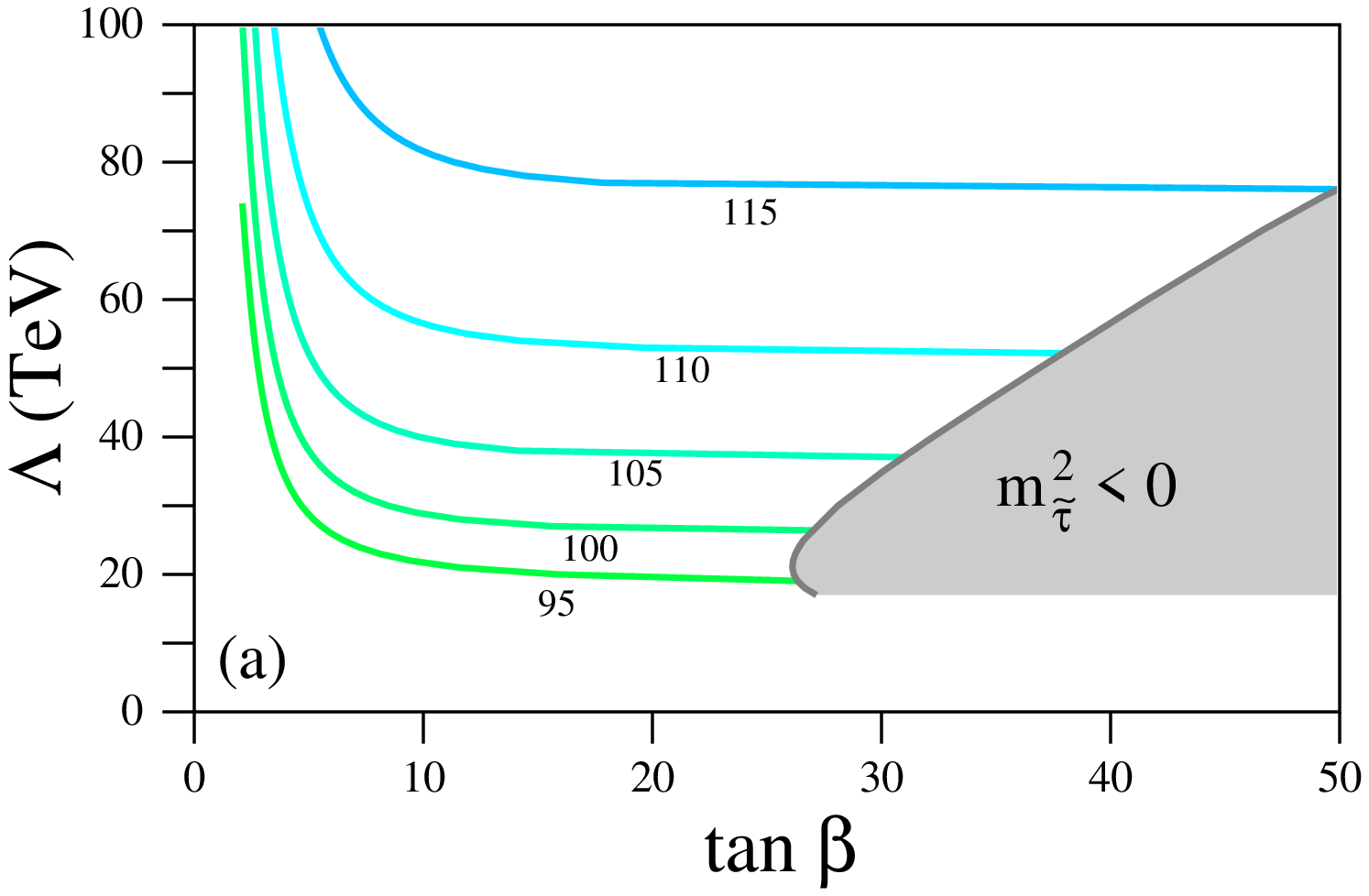}
  \epsfbox[0 0 504 300]{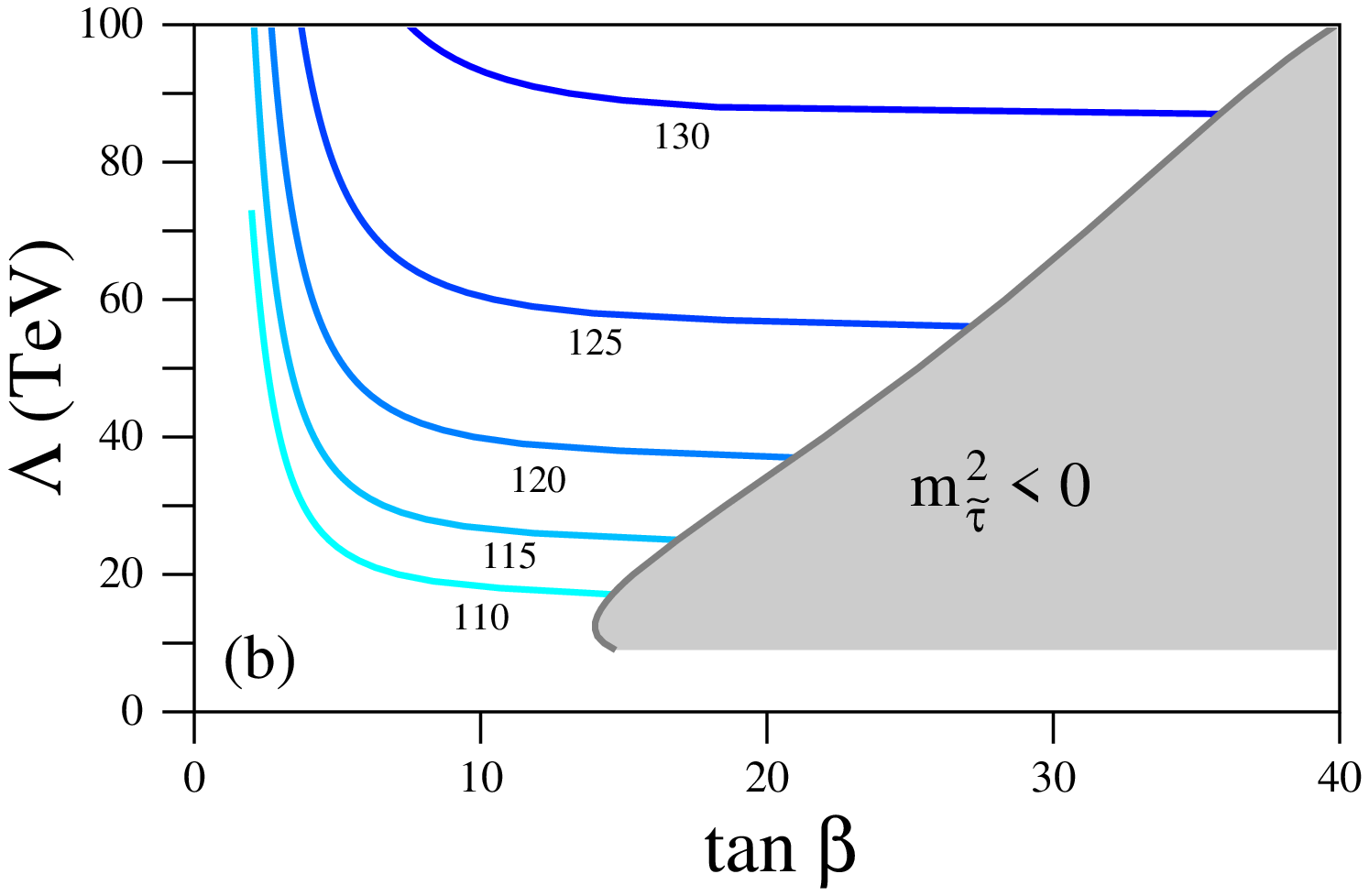}
\end{figure}

\vfil\eject

\end{document}